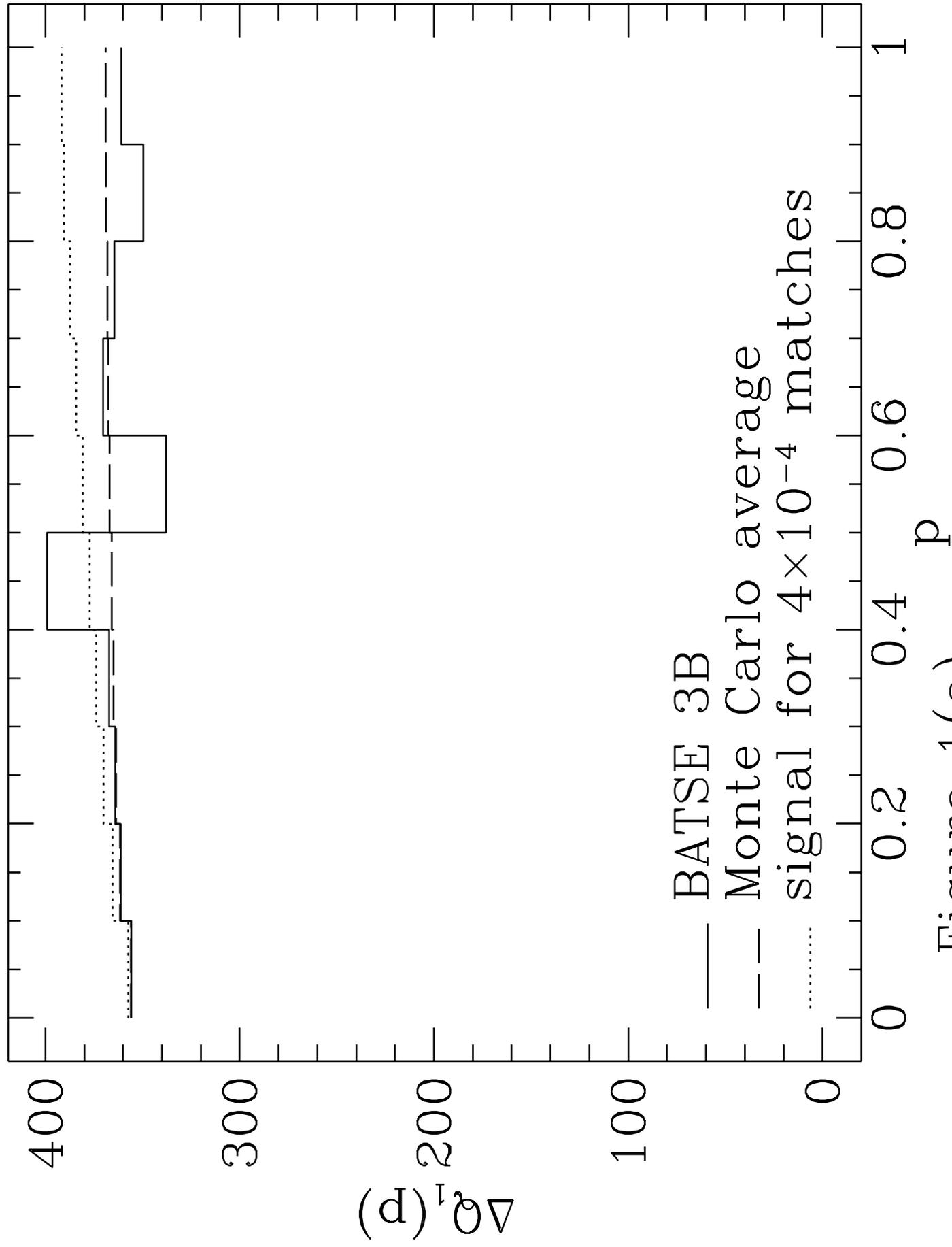

Figure 1(a)

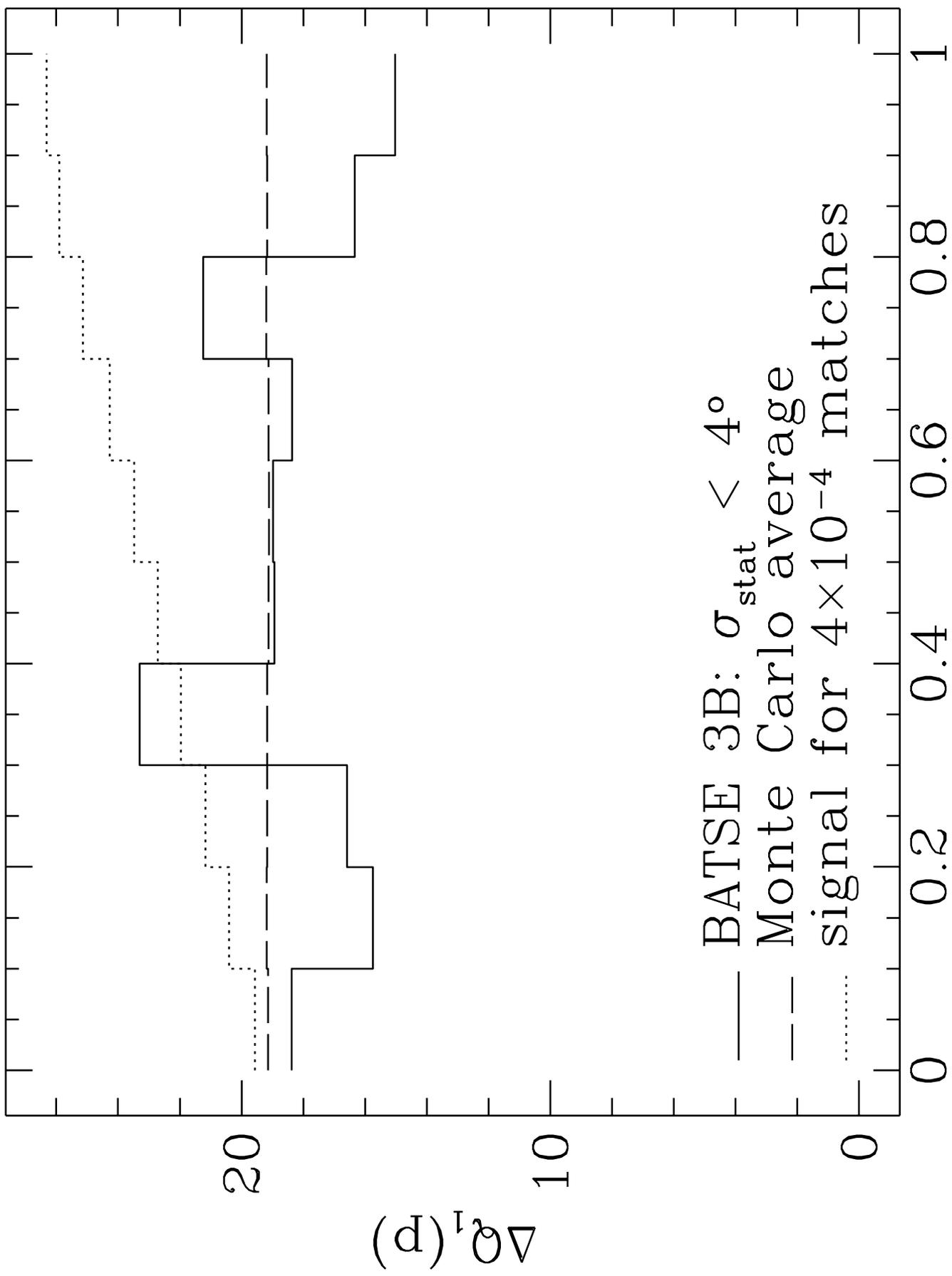

Figure 1(b)

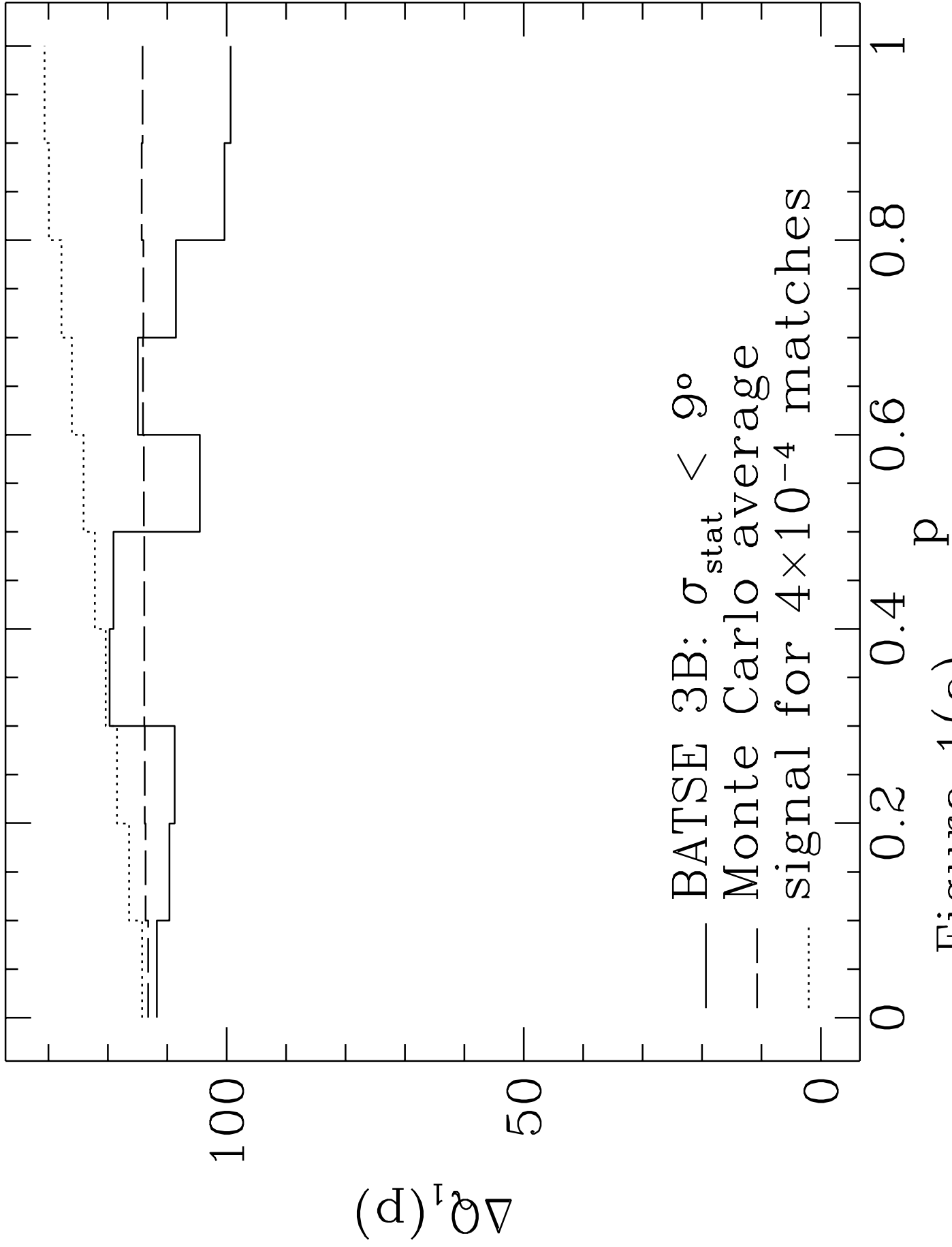

Figure 1(c)

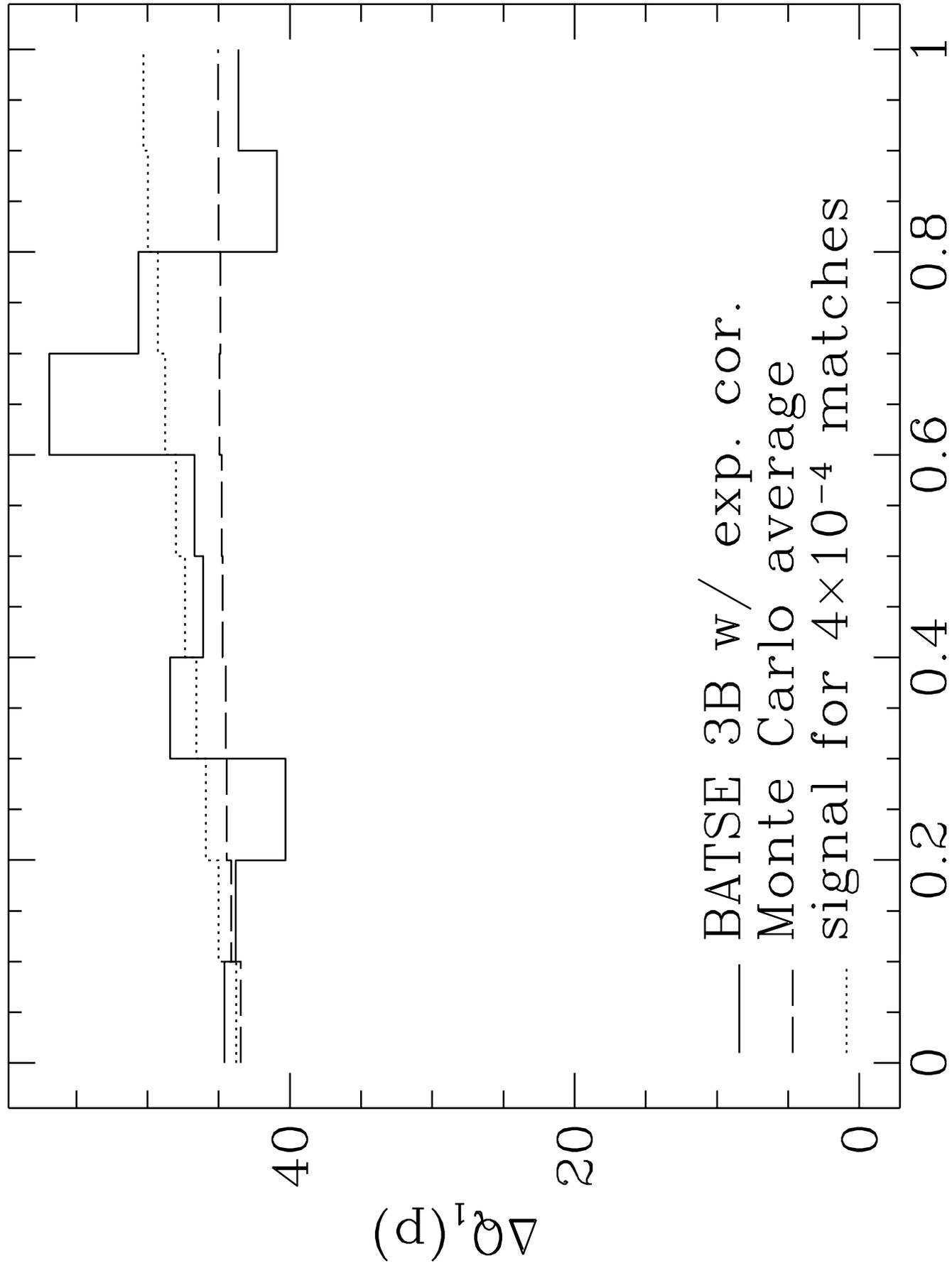

Figure 1(d)

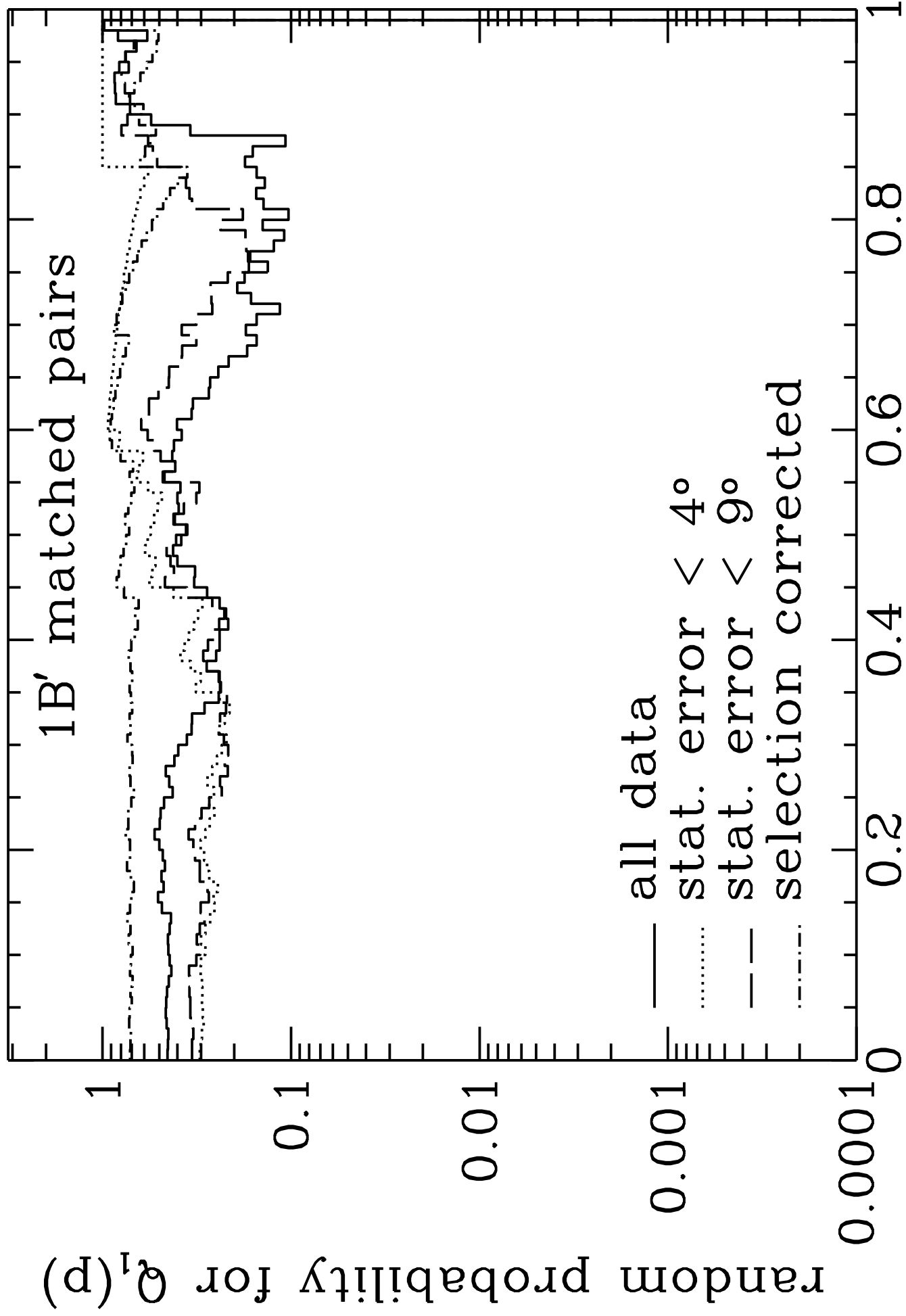

Figure 2(a)

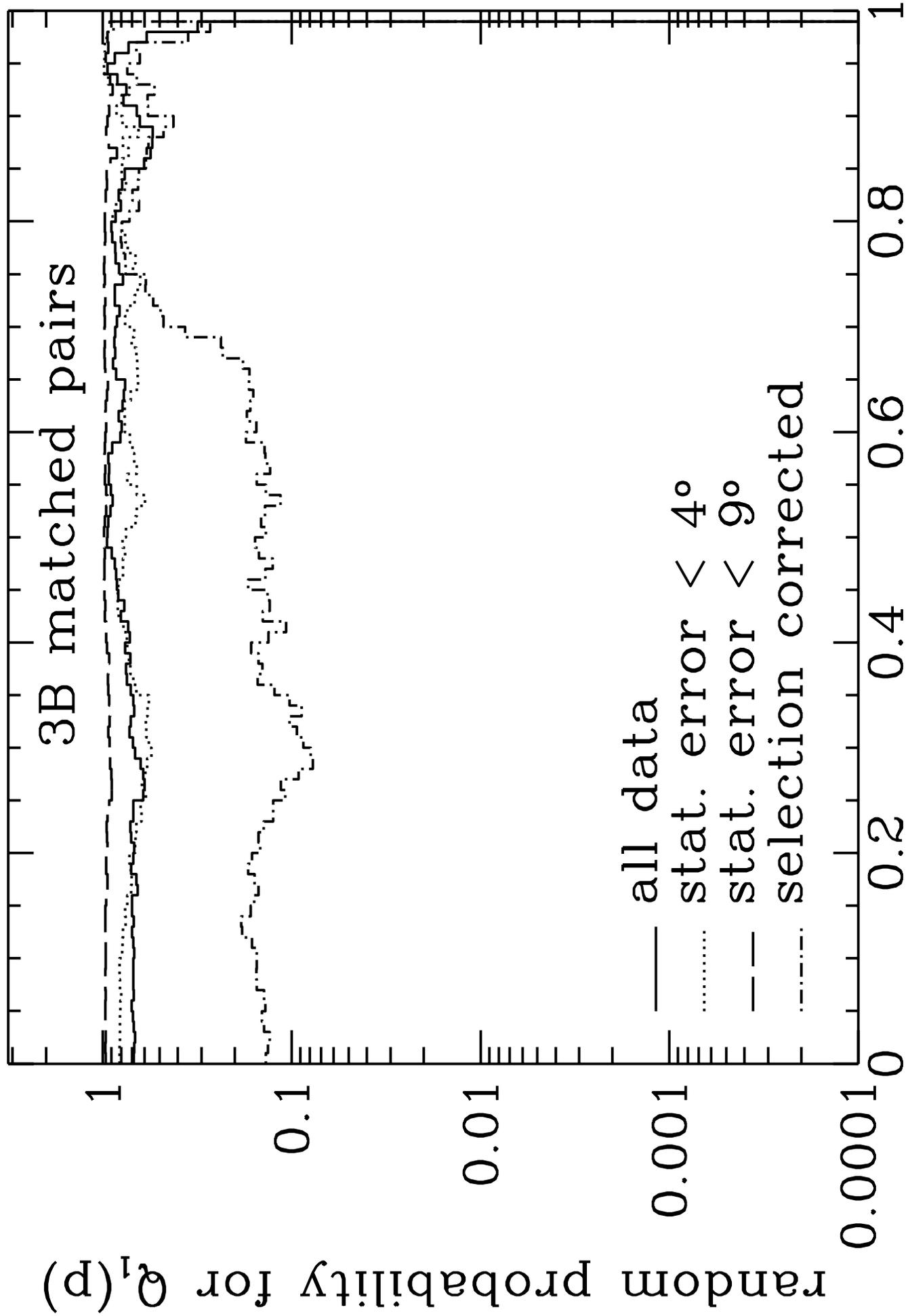

Figure 2(b)

# A Test of Gamma Ray Burst Recurrence in the BATSE 3B Data Set

DAVID P. BENNETT AND SUN HONG RHIE

*Center for Particle Astrophysics, University
of California, Berkeley, CA 94720*

and

*Institute of Geophysics and Planetary Physics, Lawrence
Livermore National Laboratory, Livermore, CA 94550*

## ABSTRACT

We analyse the BATSE 3B catalog using the pair-matching statistic. This statistic counts only the burst pairs which may have originated from the same source, so it is less likely to yield false detections of "repeating bursts" than the nearest neighbor and correlation function statistics. Even in the ideal case when repeating is the only possible source of burst correlations, the pair matching statistic is more sensitive to repeating bursts than these other statistics particularly for models which predict faint or multiple burst repetitions. We find that the BATSE 3B data set contains no excess of matched burst pairs over the expectation from a sample with random positions. We also apply the pair-matching statistic to the bursts that previously appeared in the BATSE 1B catalog which now have improved positions and position errors in the BATSE 3B data set. Previously, these bursts had exhibited some peculiar position correlations that were interpreted by some as evidence for burst repetition, but we find that these correlations have disappeared with the improved BATSE 3B positions.

# 1. Introduction

Whether classical gamma ray bursts might recur from the same sources on the sky is an important issue because it has direct bearing on the nature of the burst sources: Do the gamma ray bursts originate from from inside the Galaxy or from cosmological distances? In cosmological models, the bursts appear to be distributed uniformly on the sky and the observer appears to be at the center of a distribution which has a dearth of faint (and presumably distant) bursts due to cosmological redshift effects. These predictions are consistent with the burst distribution observed by BATSE (Meegan, *et al.*, 1992; Fishman, *et al.*, 1993, Meegan, *et al.*, 1995a). On the other hand, the discovery of repeating bursts would be an evidence against cosmological models because they generally demand an energy output so large that the source must be consumed during the burst. Clearly, such a source can not repeat.

The implications of the publicly released BATSE 1B data set regarding burst repetition have been controversial. Quashnock and Lamb (1993) presented a "nearest neighbor" analysis of this data and claimed statistical evidence for repeating bursters. Other authors (Narayan and Piran, 1993; Nowak 1994; Strohmayer, Fenimore, and Miralles, 1994; Meegan, *et al.*, 1995b; and Efron and Petrosian 1995) have also analyzed the BATSE data primarily with nearest neighbor statistic or the angular correlation function. The correlation function analyses have found an excess of pairs at separations close to $0°$ as well as a similar excess at $180°$. The excess at $180°$ is difficult to explain as a physical effect, so perhaps the correlations at both $0°$ and $180°$ are due to a systematic error in the burst locations. One peculiarity noticed by many of these authors was that the "signal" in the correlation function occurred on a scale of $\lesssim 4°$ while the estimated position errors of the bursts responsible for the correlations were $\sim 8°$. Quashnock and Lamb suggested that this was the signal of a burst population with a few repeating bursts that were seen several times. It was also noted that many of the bursts responsible for this "signal" were relatively faint.

One problem that affects both the nearest neighbor statistic and the two-point angular correlation function is that these statistics really measure correlations between burst positions rather than the property that we are really after: whether each pair of bursts could have come from the same source. It is true that repeating burst sources will cause burst positions to be correlated, but there are other mechanisms that might also cause the bursts to be correlated, such as systematic position measurement errors. We need a statistic that will not only measure burst correlations but will determine whether each pair of bursts could have come from the same source. If there is a significant excess of 'matched' burst pairs, then is the separation distribution consistent with the expected distribution from a population of repeating sources?

In order to address this problem, we have devised the pair matching statistic[†] (Bennett & Rhie, 1996; Rhie & Bennett, 1995). For each pair of bursts, we define a match probability that measures the likelihood that this pair has originated from the same location on the sky. The match probabilities are then combined into our pair matching statistic which measures only those burst pairs which are consistent with being produced by the same source. Even in the idealized case in which *all* burst correlations are due to burst repetition, the pair matching statistic is more sensitive to burst repetition than the nearest neighbor and correlation function statistics particularly for faint or multiple repeating bursts. The sensitivity of the pair matching statistic to the fraction of repeating pairs improves as $1/N$ where $N$ is total number of bursts. (The noise scales as the square root of the total number of burst pairs). In contrast, the sensitivity of the nearest neighbor statistic improves much more slowly than this for large $N$.

Another important distinction is due to the fact that the pair-matching statistic makes use of the match probability for each pair of bursts. This allows us to compare the shape of the observed pair match probability distribution to the distribution predicted by repeater models. Thus, if the pair matching statistic for the BATSE data should reject the "null" hypothesis of no repeating bursters, then we can also apply a "goodness of fit" test to see if the observed signal is actually consistent with burst repetition. In fact, when we applied (Bennett & Rhie, 1996) the pair-matching statistics to the BATSE 1B and 2B data sets, we found that the observed small angle correlation was due to bursts separated by an angle smaller than expected based on the BATSE position error estimates. When all bursts pairs consistent with originating from the same source were considered, no significant signal of repeating bursts remained.

The main point of this paper is to apply our pair-matching statistic to the bursts in the BATSE 3B catalog. In section 2, we define the pair matching statistic and summarize our previous results. In section 3, we analyze the BATSE 3B with the pair matching statistic, and we also consider the BATSE 1B′ data set which consists of the BATSE 1B bursts with improved BATSE 3B positions.

---

[†] A similar statistic has been developed by Petrosian & Efron (1995) to study repeating on short timescales.

## 2. The Pair Matching Statistic

The pair matching statistic is discussed at length and compared to other statistics in our previous paper (Bennett & Rhie, 1996), so we will just summarize it here. If we let $\theta_{ij}$ be the angular distance between bursts $i$ and $j$, then the joint probability density of the pair is $e^{-\theta_{ij}^2/2\sigma_{ij}^2}/(2\pi\sigma_{ij}^2)$ where $\sigma_{ij}^2 = \sigma_{i,\text{stat}}^2 + \sigma_{i,\text{sys}}^2 + \sigma_{j,\text{stat}}^2 + \sigma_{j,\text{sys}}^2$ is the quadrature sum of the statistical and systematic position errors for bursts $i$ and $j$. The match probability for bursts $i$ and $j$ is defined to be

$$p_{ij} = \frac{1}{\sigma_{ij}^2} \int_{\theta_{ij}}^{\infty} e^{-\theta^2/2\sigma_{ij}^2} \, \theta \, d\theta = e^{-\theta_{ij}^2/2\sigma_{ij}^2} \; . \qquad (2.1)$$

which is just the formal probability that if bursts $i$ and $j$ originated from the same position on the sky, they would have a separation at least as large as that observed. Thus, for burst pairs that come from the same source, we should expect $p_{ij}$ to be uniformly distributed between 0 and 1.

One can express the expected distribution of $p_{ij}$ values as a distribution function: $g(p)$. For matched pairs, $g(p)$ is just a constant, so for $N_{\text{matched}}$ matched pairs, we have

$$g_{\text{matched}}(p) = N_{\text{matched}} \; , \qquad (2.2)$$

while for a set of $N_{\text{random}}$ randomly distributed burst pairs we have

$$g_{\text{random}}(p) = \sum_{i,j<i} \frac{\sigma_{ij}^2}{2p} = \frac{N_{\text{random}} \left\langle \sigma_{ij}^2 \right\rangle}{2p} \; , \qquad (2.3)$$

where we have used the small angle approximation to derive eq. (2.3). Note that $g_{\text{random}}(p)$ does not actually diverge for small $p$ because there is a minimum allowed value of $p$ which corresponds to $\theta = \pi$. In order to test for repeating bursts, we will need to be able to detect a signal of the form, (2.2), from the background noise of the form (2.3) with $N_{\text{random}} \gg N_{\text{matched}}$. Using eq. (2.1), we can define the pair match statistic:

$$Q_\alpha(p) = \sum_{\text{pairs}} p_{ij}^\alpha \Theta(p_{ij} - p) \; , \qquad (2.4)$$

where $\Theta$ is the familiar step function. The value of $Q_\alpha(p)$ for an uncorrelated data set depends on the number of data points and the distribution of error bars, so Monte Carlo simulations of random data sets with the observed distribution of error bars are required to assess the statistical significance of these statistics. For a gamma ray burst population that does repeat, then the expected signal is given by

$$Q_\alpha(p) - \langle Q_\alpha(p) \rangle_{\text{RANDOM}} = \frac{N_{\text{matched}}}{\alpha + 1} \left(1 - p^{\alpha+1}\right) - \frac{(1-p^\alpha)}{2\alpha} \sum_{\substack{\text{matched} \\ \text{pairs}}} \sigma_{ij}^2 , \quad (2.5)$$

where $\langle Q_\alpha(p) \rangle_{\text{RANDOM}}$ is the mean value of the Monte Carlo simulated BATSE catalogs with random positions, and $N_{\text{matched}}$ is the number of burst pairs originating from the same physical burst location on the sky. For $\alpha = 0$, $(1-p^\alpha)/\alpha$ is replaced by $-\ln p$ in the last term of eq. (2.5). In our previous paper, we compared the expected signal-to-noise ratio for all values of $p$ and $\alpha = 0$ or 1. In all cases, we found that $Q_1(0)$ was the most sensitive statistic, so this the statistic that we use determine if a repeater signal can be seen in the data.

We have also compared the sensitivity of the pair matching statistic to that of the two-point correlation function and the nearest neighbor statistics and found that the pair matching statistic is generally more sensitive to repeating bursters. In contrast, the nearest neighbor statistic is particularly insensitive to repeating in large data samples: In a sample of 1122 bursts with 240 burst pairs from the same location, the nearest neighbor statistic will only see a signal which is significant at the 60-90% confidence level while the pair matching statistic sees a signal which is significant at the 3-$\sigma$ level (assuming repeater burst position errors of 8°).

We should point out that although the value of $p$ in eqns. (2.1) and (2.4) is a calculated assuming Gaussian errors, the pair matching statistic *does not* rely upon the assumption that the errors are Gaussian distributed. To the extent that the position errors do not follow a Gaussian distribution, we can consider $p_{ij}$ to be a parameter which is *correlated with* rather than equal to the true match probability. Since we are comparing against the "null hypothesis" of no repeating burst sources, our Monte Carlo simulated BATSE catalogs use only random positions. Thus, the comparison to the random catalogs remains valid for any distribution of measured position errors.

## 3. Pair Matching Analysis of the BATSE 3B Dataset

Before we present the pair matching results for the BATSE 3B data set, let us discuss the subsets of the data that we will apply the statistic to. Previous authors (Quashnock and Lamb, 1993; Narayan and Piran, 1993) have considered sub-samples of the catalog defined by the requirement that $\sigma_{\text{stat}}$ be less than some value (9° for QL and 4° for NP). For comparison with their results and with our BATSE 2B results, we also include these sub-samples in the present analysis. Out of 1122 bursts in the BATSE 3B catalog, 956 and 628 bursts pass the $\sigma_{\text{stat}} < 9°$ and the $\sigma_{\text{stat}} < 4°$ cuts respectively. The BATSE 1B′ data set is the subset of the BATSE 3B data set which corresponds to the same time interval as the BATSE 1B data set. Of the 263 bursts in this data set, 229 bursts pass the $\sigma_{\text{stat}} < 9°$ cut and 147 bursts pass the $\sigma_{\text{stat}} < 4°$ cut. In the original analysis of the 260 BATSE 1B bursts, 202 and 133 bursts passed these two cuts. These differences are due to burst reclassification and improved positions.

With data sets containing a few hundred to a thousand bursts with position errors ranging from a few to a few tens of degrees, accidental matches between bursts from different sources will be common, and they will be a source of noise for the pair matching statistic. The number of accidental matches will depend somewhat on the average exposure of BATSE to different areas of sky. For example, if BATSE observed only the Northern hemisphere, the number of accidental matches would be increased because the observed bursts would be confined to a smaller region of sky. We correct for BATSE's non-uniform exposure by using the the exposure table provided with the BATSE data. However, these exposure tables do not completely describe the exposure distribution for the faintest bursts that BATSE can detect. For this reason, we have defined on exposure corrected sample which contains only those bursts for which the sky exposure can be accurately determined from the the BATSE exposure table. The details of this event selection are given in Bennett and Rhie (1996). There are 540 BATSE 3B bursts which pass this cut. 151 of these are in the BATSE 1B′ data set compared to 150 in the BATSE 1B data set.

Fig. 1 shows the variation of the $Q_1(p)$ statistic with $p$ for data samples from the BATSE 3B catalog. We have chosen to plot the differential version of this statistic $(i.e. \Delta Q_1(p) \equiv Q_1(p) - Q_1(p + \Delta p))$ because this makes it easier to see the $p$ dependence of $Q_1(p)$. Also plotted are the expected signal from a sample which has $4 \times 10^{-4}$ of all burst pairs originating from the same source and the Monte Carlo average of data sets with random positions. Clearly, only the exposure corrected sample might have a signal as large as $\sim 4 \times 10^{-4}$ level, and as shown in table 1, number of matched pairs for this sample is only about 1 $\sigma$ above the average of the Monte Carlo simulations.

The precise limits on the number of burst pairs that might originate from the same source are shown in Table 1. For the full sample the 2 $\sigma$ upper limit on the number of bursts that might have originated from the same sources is 120 or about $1.9 \times 10^{-4}$ of all the burst pairs. It is somewhat more conventional to quote the number of bursts that are due to repeating sources, but this is an ambiguous number for the pair matching statistic (as well as for the correlation function) because it is based on burst pairs rather than single bursts. For example, if each burst is seen to repeat only once, a fraction of $1.9 \times 10^{-4}$ repeat pairs corresponds to 240 single bursts or 21 % of the 1122 total that are from repeating sources. However, the 120 pairs could also be due to 1 source that generates 16 bursts. In this case, the fraction of all bursts due to repeaters would be 1.4%.

Since our main question is whether there is *any* statistically significant signal of burst recurrence in the same position, the fraction of simulated BATSE catalogs with random burst positions that have $Q_1(p) \geq$ the observed value is of particular interest. These values are plotted in Fig. 2 for both the BATSE 3B data set and its 1B$'$ subset for both "matches" and "antipode matches." These plots indicate that there is no significant $Q_1(p)$ signal for any of our data subsets for any value of $p$ except for perhaps the BATSE 3B antipodal pairs for $p > 0.7$ where the random probability of exceeding the observed $Q_1(p)$ value is $\sim 0.01$. Given the large number of data subsets and the range of $p$ values we've examined, this can hardly be considered significant, however.

It is of particular interest to note that the BATSE 1B$'$ data set does not show any hint of a matched pair or antipode pair signal. This differs from the results for the original release of the BATSE 1B data set. In the original 1B data set, the "match" statistics for the $\sigma_{stat} < 9°$ sample (and to a lesser extent in some of the other samples), there appeared to be some sort of a signal in the measured $Q_1(p)$ values for $p \gtrsim 0.6$. The random probability of the observed values was a few $\times 10^{-4}$, but for smaller $p$ values where the signal to noise should have been larger the random probability of the observed value rose to reach about 2% at $p = 0$. Thus, there appeared to be a significant excess of burst pairs separated by angles smaller than expected for repeating bursts and a much more marginal signal that might have been consistent with a small population of repeating bursts. Undoubtedly, burst pair excess at $p \gtrsim 0.6$ is the same signal that had been seen by other authors using the nearest neighbor and correlation function statistics. Thus, the refined burst position analysis used for the BATSE 3B data release has clearly removed the position correlations that some had interpreted as a signal of repeating bursts. The peculiar correlation at 180 degrees has also disappeared. This suggests that both the small angle and the antipodal clustering seen in the original BATSE 1B data set was due to systematic effects associated with the method of determining burst positions.

In summary, we have used our pair matching statistic to search for evidence of burst repetition in the BATSE 3B data set, and we find no evidence for burst repetition at a level of sensitivity that is superior to the sensitivity that can be attained with the nearest neighbor statistic or the correlation function. We have also analyzed the BATSE 1B$'$ subset of the BATSE 3B data set and have found that the position correlations that existed in the 1B data set have now disappeared with the improved BATSE 3B position measurements. This suggests that the correlations in the BATSE 1B data might have been caused by systematic errors in the original position determinations that have now been corrected.

## ACKNOWLEDGEMENTS

This work was supported in part by the National Science Foundation through the Center for Particle Astrophysics and by the U.S. Department of Energy at the Lawrence Livermore National Laboratory under contract No. W-7405-Eng-48.

| Sample: | All | $\sigma_{\text{stat}} < 9°$ | $\sigma_{\text{stat}} < 4°$ | Exposure corrected |
|---|---|---|---|---|
| $N_{\text{match}}$ 3B | $-43 \pm 82$ | $-85 \pm 47$ | $-17 \pm 20$ | $32 \pm 29$ |
| $N_{\text{antipode}}$ 3B | $25 \pm 82$ | $59 \pm 47$ | $19 \pm 20$ | $-18 \pm 29$ |
| match significance 3B | 68.55% | 96.74% | 81.50% | 13.74% |
| antipode significance 3B | 36.40% | 10.51% | 16.35% | 72.35% |
| $N_{\text{match}}$ 1B′ | $2 \pm 19$ | $5 \pm 12$ | $2 \pm 5$ | $4 \pm 7$ |
| $N_{\text{antipode}}$ 1B′ | $-8 \pm 19$ | $10 \pm 12$ | $4 \pm 5$ | $9 \pm 7$ |
| match significance 1B′ | 45.15% | 33.61% | 30.52% | 71.45% |
| antipode significance 1B′ | 65.57% | 19.17% | 20.83% | 9.51% |
| $N_{\text{match}}$ 1B | $13 \pm 25$ | $27 \pm 14$ | $1 \pm 7$ | $3 \pm 12$ |
| $N_{\text{antipode}}$ 1B | $54 \pm 25$ | $24 \pm 14$ | $17 \pm 7$ | $9 \pm 12$ |
| match significance 1B | 29.3 % | 2.65% | 43.36% | 36.96% |
| antipode significance 1B | 2.42% | 4.54% | 1.32% | 21.92% |

*Table 1.* The number of matches and anti-matches is given for the BATSE 3B, 1B′, and 1B data sets as measured with the $Q_1(0)$ statistic with 1-$\sigma$ error bars determined from the RMS deviation from the mean of 10,000 simulated BATSE catalogs. The significance refers to the fraction of the 10,000 simulated catalogs with larger values of $Q_1(0)$ than the real data. The 1B′ data set is the subset of the 3B catalog that corresponds to the 1B catalog.

# FIGURE CAPTIONS

1. $\Delta Q_1(p) \equiv Q_1(p) - Q_1(p + \Delta p)$ for the BATSE 3B catalog is plotted as a function of $p$ for all bursts, (a), bursts with $\sigma_{\text{stat}} < 9°$, (b), bursts with $\sigma_{\text{stat}} < 4°$, (c), and bursts which pass the "exposure correction" cut, (d). $\Delta p = 0.1$ is the histogram bin size. $Q_1(p)$ is obtained from $\Delta Q_1(p')$ by summing over all $p' > p$.
2. The fraction of 10,000 simulated BATSE datasets which have $Q_1(p)$ values which exceed $Q_1(p)$ for the real data is plotted as a function of $p$ for (a) the BATSE 1B$'$ subset of the BATSE 3B catalog, and (b) the BATSE 3B catalog.